# Scan-specific, Parameter-free Artifact Reduction in K-space (SPARK)

Onur Beker[1,2], Congyu Liao[1,3], Jaejin Cho[1,3], Zijing Zhang[1,4], Kawin Setsompop[1,3,5], and Berkin Bilgic[1,3,5]

[1]Martinos Center for Biomedical Imaging, Charlestown, MA, United States, [2]Electrical and Electronics Engineering, Bogazici University, Istanbul, Turkey, [3]Harvard Medical School, Boston, MA, United States, [4]College of Optical Science and Engineering, Zhejiang University, Hangzhou, China, [5]Harvard/MIT Health Sciences and Technology, Cambridge, MA, United States

*This is a reformatted version of the original abstract. To see the original (with expandable figures), please use the following link to the submission site: https://submissions2.mirasmart.com/ISMRM2020/ViewSubmissionPublic.aspx?sei=H3og0nQSO*

## Synopsis

**We propose a convolutional neural network (CNN) approach that works synergistically with physics-based reconstruction methods to reduce artifacts in accelerated MRI. Given reconstructed coil k-spaces, our network predicts a k-space correction term for each coil. This is done by matching the difference between the acquired autocalibration lines and their erroneous reconstructions, and generalizing this error term over the entire k-space. Application of this approach on existing reconstruction methods show that SPARK suppresses reconstruction artifacts at high acceleration, while preserving and improving on detail in moderate acceleration rates where existing reconstruction algorithms already perform well; indicating robustness.**

## Introduction

Parallel imaging reconstruction for accelerated acquisitions is generally posed as an optimization problem, where a data consistency term is used to match the sampled points, and regularization terms may be added to impose prior knowledge about image characteristics [1-5]. One line of work is GRAPPA and its variants [1-4], where convolutional kernels are used to interpolate missing k-space points. These kernels are calibrated to enforce linear dependencies in the fully sampled autocalibration signal (ACS). This ACS region can be reweighted to better emphasize high frequency regions [2]. Image phase can also be utilized to provide additional encoding via the virtual coil (VC) concept [3,4]. Another approach is to use neural networks for non-linear interpolation [6-7]. One such method is RAKI, where subject-specific CNNs trained on ACS data are used to interpolate missing k-space points [6]. Limited degrees of freedom in receive arrays prevent parallel imaging from achieving high acceleration rates. Nonlinear interpolation fails to address this from limited ACS data as parameter estimation becomes ill-conditioned. We propose SPARK to lift these barriers by estimating reconstruction artifacts. SPARK assumes that these artifacts can be captured from the k-space errors in the ACS alone in a way that generalizes to the entire k-space, much like how dependencies captured from the ACS can be used to reconstruct missing points across the entire k-space. After these artifacts are estimated, they can be corrected for.
Code/data: *github.com/bekeronur/SPARK*.

## Methods

Training and artifact correction are depicted in Figure 1. After undersampled coil k-space is reconstructed with an existing method (e.g. GRAPPA, LORAKS), an individual CNN is trained to predict a correction term for the reconstructed k-space of that particular coil (i.e. $2n_c$ networks are trained in total). Each of these coil CNNs takes as input all of the reconstructed k-spaces across all coils, resulting in $2n_c$ input channels ($n_c$ for real/imaginary parts). The network architecture was determined empirically. Each SPARK network $\mathbf{S_c}$ for coil $c$ is trained with the following objective:

$$\min_{\mathbf{w_c}} \| crop_{ACS}[ S_c(\mathbf{w_c}; k) ] - d_c \|_2$$

where $w_c$ are the kernel weights, $k$ are the reconstructed k-spaces across all coils. $d_c = ACS_c - crop_{ACS}[k_c]$ is the difference between the ACS lines of coil $c$ and the corresponding lines in the reconstructed k-space for the same coil (where reconstruction is performed without substituting the ACS back). The predicted correction term is then added to the initial reconstruction.

To demonstrate that SPARK can work synergistically with advanced reconstruction methods, we propose a refined GRAPPA employing Sparsity and VC concept (SVC-GRAPPA). This multiplies undersampled k-space data with diagonal matrices that implement horizontal and vertical gradient operators. VC-GRAPPA is then applied on the sparse gradient images, which are finally combined with a least squares formulation where data consistency is enforced [8]. Exploiting sparsity allows SVC-GRAPPA to outperform GRAPPA especially at high acceleration.

**Imaging Parameters:**

**Figure 2:** a volunteer was scanned with a 3T Siemens Prisma using a fully sampled GRE with FOV = 220x220mm$^2$, in-plane resolution = 0.7x0.7mm$^2$, matrix size = 320x320, slice thickness = 4mm, TR/TE = 500ms/14ms, flip angle = 70º, 27 slices, bandwidth = 360Hz/pixel. Images were retrospectively undersampled to R = {4,5,6}. ACS size for all reconstructions is 40x320.

**Figure 3:** a volunteer was scanned with a Siemens 3T Skyra using a fully sampled MPRAGE at 1mm$^3$ with FOV = 234x188x192mm$^3$. Parameters were: TR=2530ms, TI=1100ms, TE=1.7ms, flip angle = 7º, and bandwidth = 651Hz/pixel. Images were retrospectively undersampled to R = {5,6,7}. The ACS size used for all reconstructions is 30x188.

Both experiments used a 32-channel head-coil for reception. ACS data are substituted back in the conventional reconstructions.

# Results

Figures 2 and 3 show the performance of SPARK applied on GRAPPA reconstructions. These results are compared with optimized (in terms of kernel size and Tikhonov regularization) GRAPPA and standard RAKI. In Figure 2, SPARK decreases RMSE by up to **2.1-fold** compared with GRAPPA, and by up to **1.3-fold** compared with RAKI. In Figure 3, SPARK reduces RMSE by up to **2.1-fold** over GRAPPA; and by up to **1.7-fold** over RAKI. SPARK's ability to mitigate artifacts can be observed in the difference images and in k-space. Figure 4 compares SPARK, RAKI and GRAPPA for different ACS sizes and accelerations. SPARK is again applied to an initial GRAPPA reconstruction. Figure 5 shows the application of SPARK to advanced initial reconstructions, where up to **1.5**- and **1.2-fold** RMSE gains were observed over SVC-GRAPPA and LORAKS.

# Discussion

We have proposed SPARK to combine sensitivity encoding and CNNs to enable high acceleration rates with improved quality. This network is trained in a scan-specific manner, only on the difference between ACS data and the corresponding region of a reconstructed k-space. This allows our method to work synergistically with well-established reconstruction methods without introducing artificial features. Beyond its ability to reduce RMSE and improve image quality at high acceleration, SPARK continues to provide benefits at moderate acceleration, where the initial reconstruction has minor artifacts. SPARK is also able to operate with stringent ACS sizes, where data-hungry nonlinear models fail to improve over GRAPPA. These properties suggest that SPARK is robust; with resulting images being at least as good as the initial reconstruction.

# Acknowledgements

This work was supported by NIH research grants: R01EB020613, R01EB019437, R24MH106096, P41EB015896, shared instrumentation grants: S10RR023401, S10RR019307, S10RR019254, S10RR023043 and NVIDIA GPU grants.

# Figures

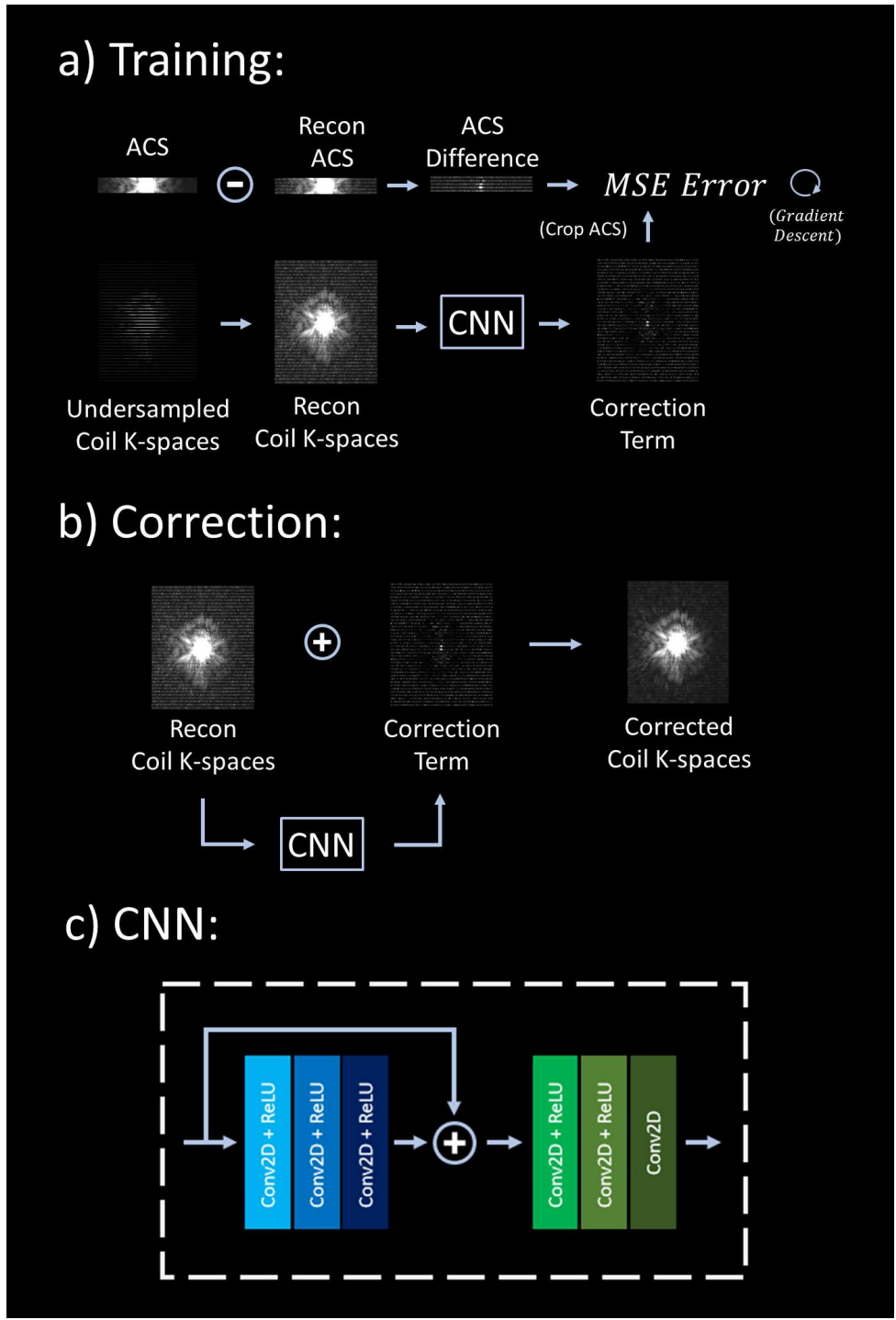

*Figure 1.* Flowchart for SPARK. Undersampled coil k-spaces are reconstructed using a conventional method, and corrected using a CNN that is trained only on scan-specific data. (a) The reconstructed coil k-spaces are passed through a CNN to obtain an estimate of the reconstruction error for each coil. This CNN is trained using the difference between the ACS lines and the corresponding lines of the reconstructed k-space. (b) After training, the estimated difference is added as a correction term to the reconstructed k-space. (c) The CNN architecture consists of a residual block, followed by a head part that reduces the filter depth progressively to one.

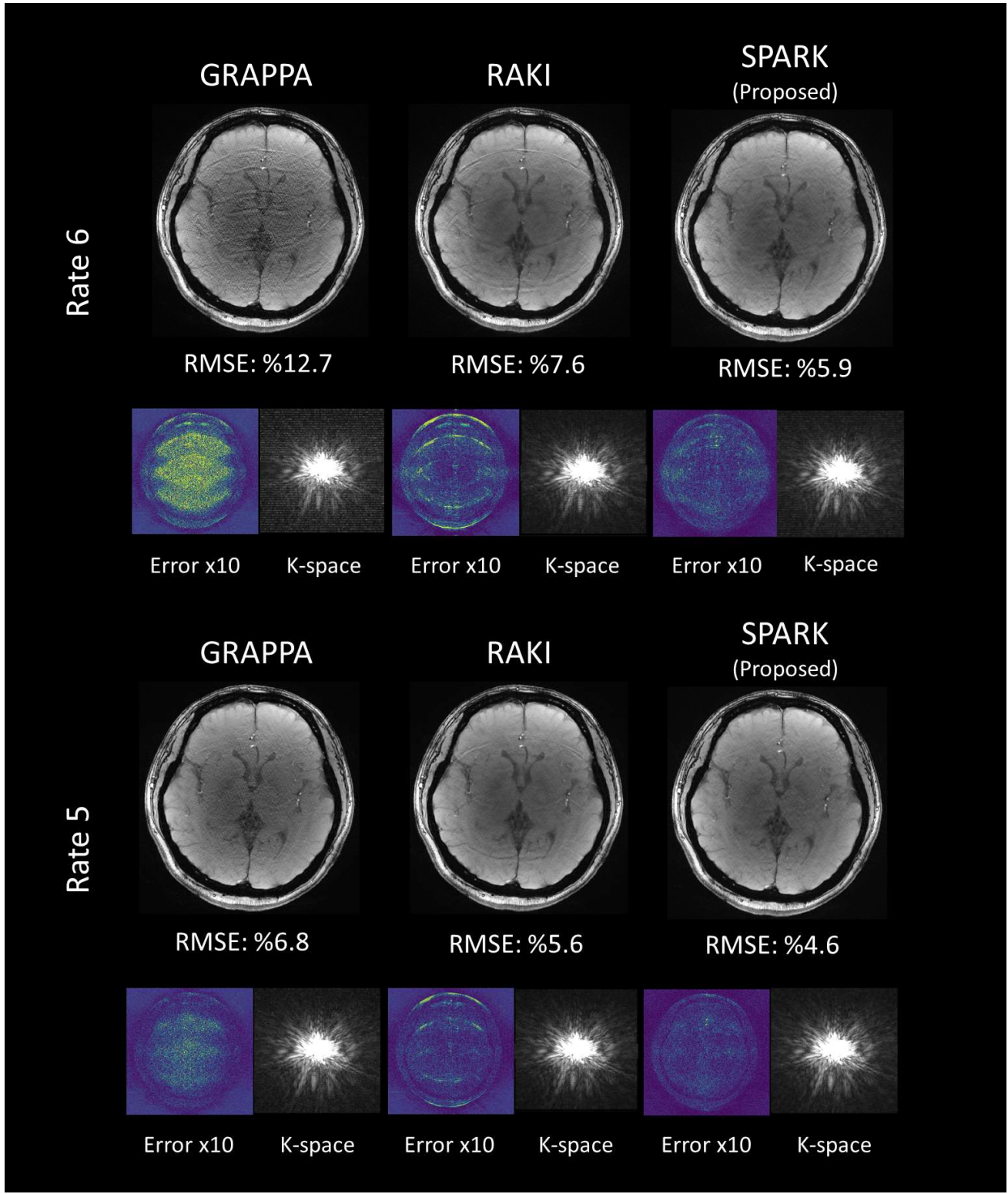

Figure 2. SPARK (applied to an initial GRAPPA reconstruction) compared with GRAPPA and RAKI on 3T GRE data, with R = 5, 6. When the acceleration rate is high and the ACS size is small, GRAPPA suffers from artifacts and noise amplification, and RAKI contains edge artifacts because it doesn't have enough data. SPARK manages to reduce both noise and edge artifacts that are present in the initial GRAPPA reconstruction that it was applied on.

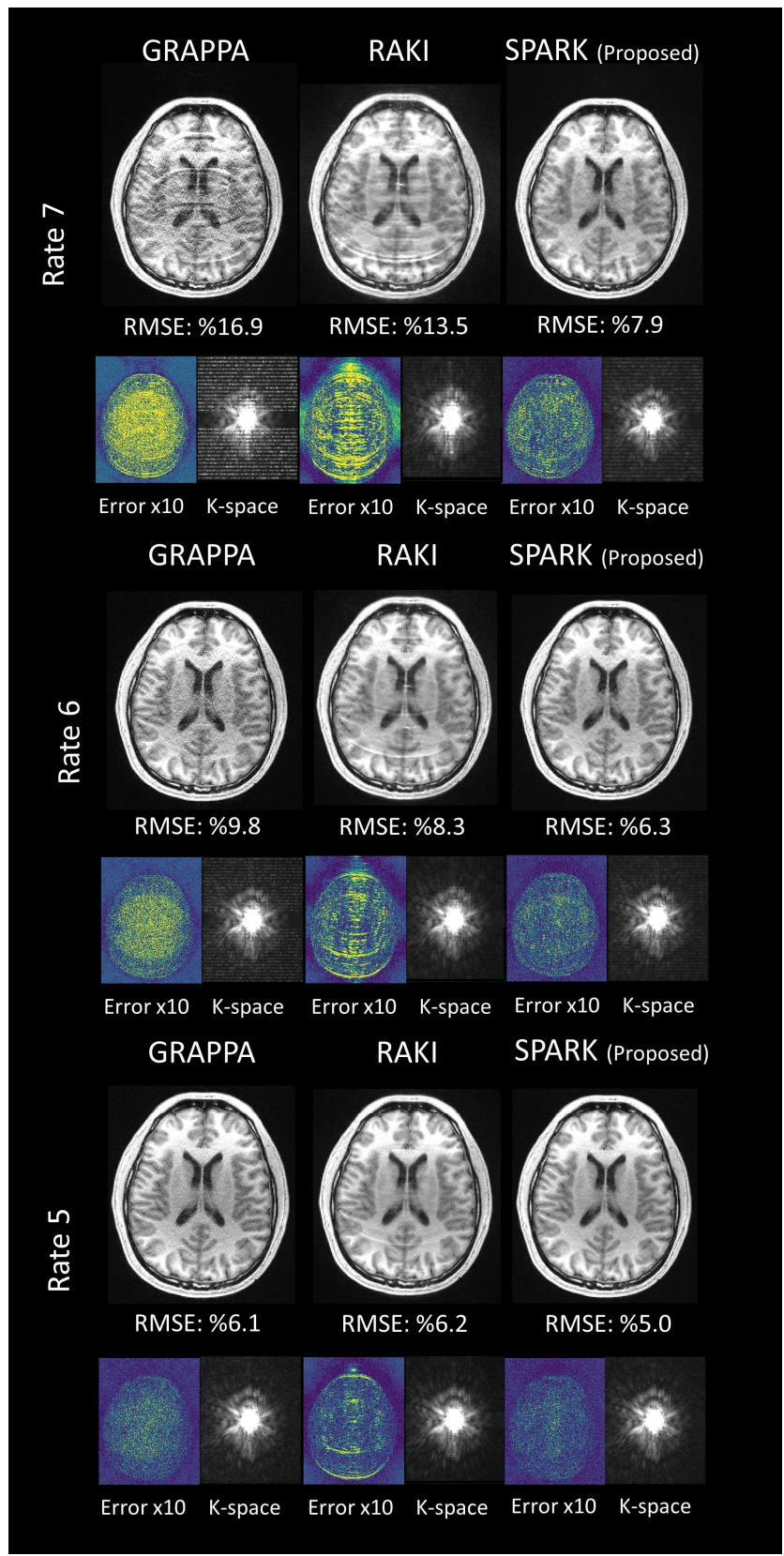

Figure 3. Comparison of SPARK, RAKI and GRAPPA for 3T MPRAGE data, with R = 5, 6, 7. SPARK manages to suppress the edge artifacts at R = 7, and continues to provide benefits at R = 5 where the initial GRAPPA reconstruction only has minor artifacts.

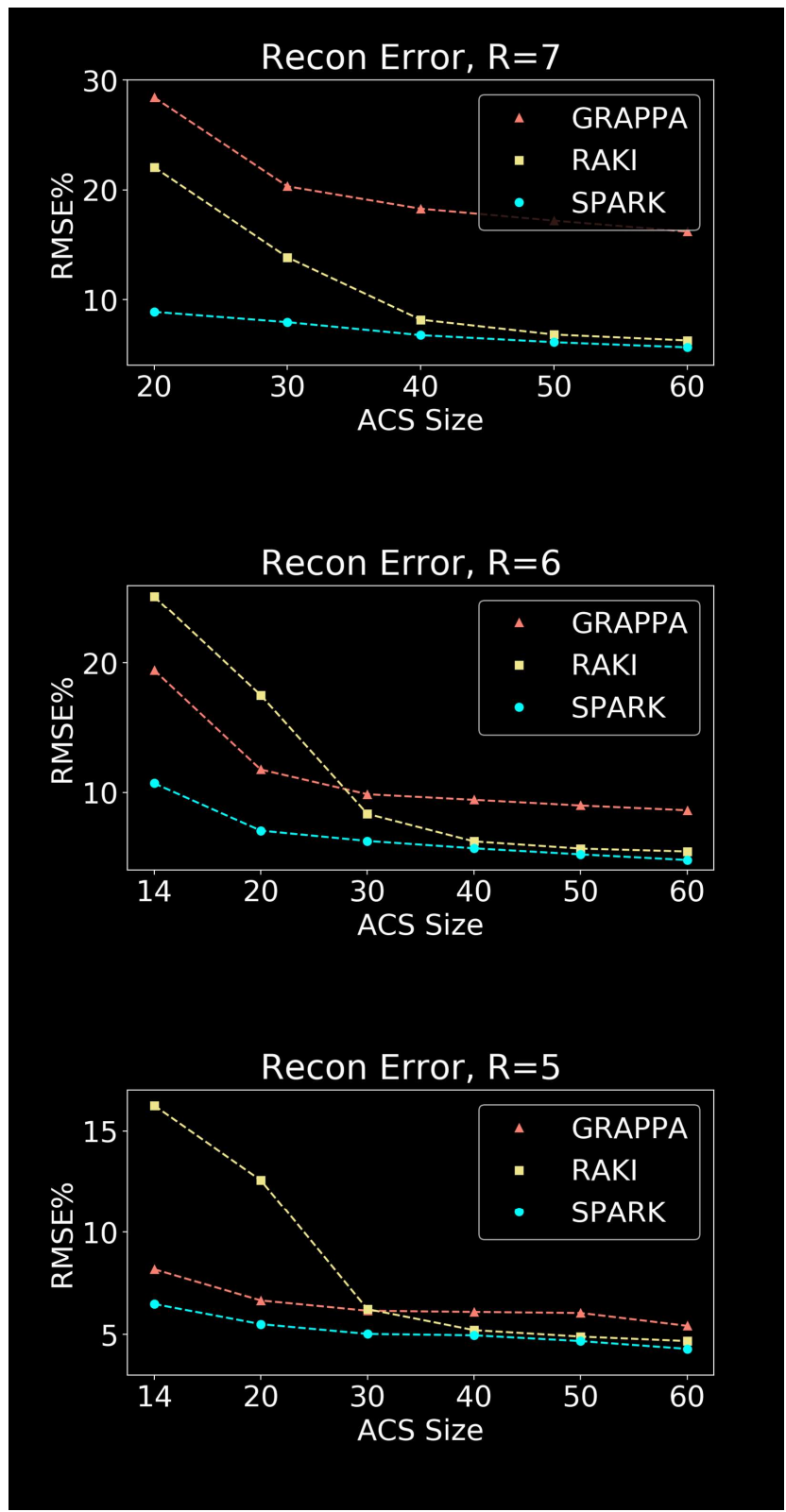

*Figure 4.* Comparison of SPARK, RAKI and GRAPPA for different ACS sizes, with R = 5, 6, 7. RAKI breaks down for small ACS sizes, since it has more parameters to estimate from limited ACS data, compared to GRAPPA. SPARK still works when the ACS size is small, because it only estimates a residual correction term, and doesn't need to learn how to reconstruct the k-space.

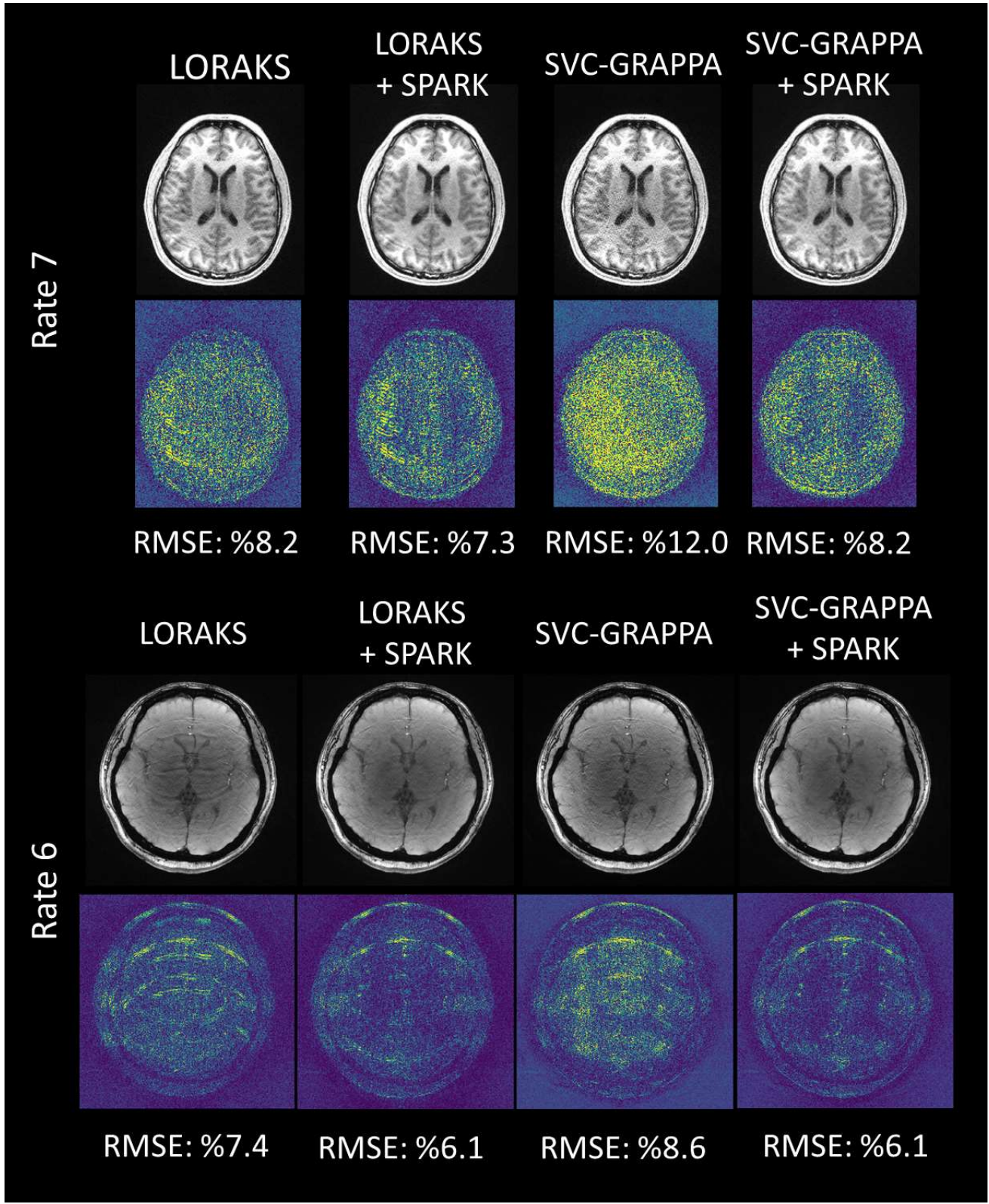

*Figure 5.* Application of SPARK to other initial reconstructions. SPARK mitigates both noise and edge artifacts present in LORAKS and SVC-GRAPPA, demonstrating that it can work synergistically with conventional reconstruction methods.